\documentclass[12pt]{iopart}
\usepackage{iopams}
\usepackage{graphicx}
\usepackage{subfigure}

\newcommand{\ii}{{\rm i}}
\newcommand{\ee}{{\rm e}}

\begin{document}
\title{Asymptotic eigenvalue distribution of large Toeplitz matrices}

\author{Seung-Yeop Lee$^1$, Hui Dai$^1$ and Eldad Bettelheim$^{1,2}$}

\address{$^1$ James Franck Institute, University of Chicago, 5640 South Ellis Avenue, Chicago, Illinois 60637, USA}
\address{$^2$ The Racah Institute of Physics, The Hebrew University of Jerusalem,
Safra Campus, Givat-Ram, 91904, Israel}
\eads{\mailto{duxlee@uchicago.edu}, \mailto{hdai@uchicago.edu}, \mailto{eldad@uchicago.edu}}
\begin{abstract}
We study the asymptotic eigenvalue distribution of Toeplitz matrices generated by a singular symbol.
It has been conjectured by Widom that, for a generic symbol, the eigenvalues converge to the image of the symbol.
In this paper we ask how the eigenvalues converge to the image.
For a given Toeplitz matrix $T_n(a)$ of size $n$, we take the standard approach of looking at $\det(\zeta-T_n(a))$, of which the asymptotic information is given by the Fisher-Hartwig theorem.
For a symbol with single jump, we obtain the distribution of eigenvalues as an expansion involving $1/n$ and $\log n/n$.
To demonstrate the validity of our result we compare our result against the numerics using a pure Fisher-Hartwig symbol.
\end{abstract}

\vspace{2pc}
 
\submitto{\JPA}

\section{Introduction}
Given a basis $\{e_1,e_2,...\}$ in a Hilbert space, a linear operator $A$ is represented by an infinite matrix whose element is given by $a_{jk}=\langle e_j,Ae_k\rangle$.  A fundamental question that is obviously important in practical application is one about approximating the linear operator by finite matrices \cite{Morrison}.

The study of Toeplitz system can be motivated by the same question, only being specified to the following situation.  We take $\{\ee^{\ii k p}|k\in{\bf Z}\}$ as the basis for the functions defined on the unit circle ${\bf T}=\{\ee^{\ii p}|p\in[-\pi,\pi)\}$.  We also assume the linear operator $A$ to be a multiplication operator, i.e.  $(A\psi)(t)=a(t)\psi(t)$ with $t=\ee^{\ii p}\in{\bf T}$.
Then the operator $A$ is represented by the infinite matrix,
$$A_{jk}=a_{j-k}:=\int_{-\pi}^{\pi}\frac{dp}{2\pi}\ee^{-\ii(j-k)p}a(\ee^{\ii
p}).$$
The semi-infinite part ($j$ and $k$ is non-negative integer) of such matrix whose $(j,k)$th component is
given by $a_{j-k}$, is called {\it Toeplitz matrix} \cite{book, book2}.

The generating function $a(\ee^{\ii p})$ is called the {\it symbol} of
the Toeplitz Matrix, and it is a function from the unit circle ${\bf T}$ to complex number ${\bf C}$.  The $n\times n$ Toeplitz matrix generated from the symbol $a$ is denoted by $T_n(a)$.
We will call $a({\bf T}):=\{a(\ee^{\ii p})|p\in[-\pi,\pi)\}$ the {\it image of the symbol}.

Toeplitz matrices are ubiquitous in physics and mathematics (\cite{Korepin, Ehrhardt, Forrester}, just to name a few applications).  In fact, the early development in the field arose from works on the two dimensional Ising model \cite{MPW,Wu,Kadanoff} where the spin-spin correlation function is written as a Toeplitz determinant.

In all these applications, the asymptotics of Toeplitz determinants feature prominently.  They are in many cases given by Szeg\"o's theorem \cite{Szego} or by the Fisher-Hartwig theorem \cite{FH,Eh}.  The former applies to smooth symbols whereas the latter contains singularities such as jumps and zeros.  The case with singularities is much more complicated, is not fully understood, and still has open questions.\footnote{The Fisher-Hartwig theorem is promoted to a theorem from the conjecture by the works of many people including Widom, Basor, B\"ottcher, Silberman, Libby, and Ehrhardt.  There is also a refined version of the conjecture by Basor and Tracy \cite{BT}. References can be found in \cite{book}.}

One of these questions (which was also raised in \cite{BM}) is: {\it Given a set of $n$ by $n$ Toeplitz matrices, what can we say about the asymptotic eigenvalue distribution for large $n$?}  Since a Toeplitz matrix comes from a multiplication operator $a(t)$ one may guess that the eigenvalues approximate the spectrum of the multiplication operator, which is simply the image of the symbol.
Or, one may expect differently since the spectrum of an infinite Toeplitz matrix consists of the smallest convex set containing the image of the symbol.
It was conjectured by Widom \cite{Widom} that, except in rare cases, the eigenvalues approximate the image of the symbol as $n$ grows.
For instance, symbols containing a single jump singularity exhibits such behavior of eigenvalues \cite{Widom}.  According to an introductory review article \cite{BM} multiple singularities may lead to an interesting phenomena involving ``stray eigenvalues", which we do not consider in this paper.

In this paper, we consider the eigenvalue distributions that approximate the image of the symbol.  We ask {\it how} the eigenvalues approaches the image as $n$ grows, i.e. {\it what} the deviation of the eigenvalues from the image is.\footnote{Fundamentally the same question has been considered by B\"ottcher, Embree and Trefethen  \cite{slow} using the {\it pseudospectra} \cite{pseudo,pseudobook}.
Instead of directly looking at the spectrum, they analyzed the resolvent and found some interesting asymptotic behaviors of the pseudospectra for a pure Fisher-Hartwig symbol.}

Let us briefly describe our method.  An eigenvalue $\lambda$ of $T_n(a)$, an $n\times n$ Toeplitz matrix, is a solution of $\det[\zeta-T_n(a)]=0$.   Therefore, the asymptotic information about the eigenvalues can be obtained from the asymptotic information of $\det[\zeta-T_n(a)]$.  This determinant is a standard object in studying the eigenvalues and has been used by many including Widom.
By applying the Fisher-Hartwig theorem, we will see that this determinant, as a function of $\zeta$, has a line of discontinuity at the image of the symbol.  This discontinuity, then, describes the deviation of eigenvalues from the image.
We note that our result concerns only the eigenvalues that are near the image of the symbol and cannot tell much about isolated eigenvalues, or stray eigenvalues.

The paper is organized as follows. In section 2, we explain Szeg\"o's theorem and the Fisher-Hartwig theorem.  We then define the spectral measure and pose our problem of finding the asymptotic spectral measure.  We explicitly calculate, for symbols with a single jump singularity, the asymptotic spectral measure in large $n$ expansion.
In section 3, we show some plots that compare our results against the numerically evaluated eigenvalues.

\section{Asymptotic eigenvalue distribution\label{section-theorems}}
\subsection{Szeg\"o's theorem and Fisher-Hartwig theorem}

Here we explain Szeg\"o's theorem and its generalization, the Fisher-Hartwig theorem.  They describe the asymptotic behavior of the determinant for a broad class of Toeplitz matrices.

Recalling that the symbol $a$ is a function from ${\bf T}$ to complex number ${\bf C}$, we define the {\it winding} of the symbol as:
\begin{equation}
\mbox{wind}\,(a)=\frac{1}{2\pi\ii}\int_{-\pi}^{\pi}d\log a(\ee^{\ii p}),\end{equation}
where the branch of log can is taken such that it does not cross the image of the symbol.

{\it Szeg\"o's (strong limit) theorem} states as follows.
If $a$ has no zeros on ${\bf T}$ and $\mbox{wind}\,(a)=0$, then
\begin{equation}\label{szego}
\lim_{n\rightarrow\infty}\frac{\det T_n(a)}{\exp[{n H(a)]}
}=\exp[{E(a)}].
\end{equation}
The two constants $H(a)$ and $E(a)$ are given respectively by
\begin{equation} \label{GofA}
H(a)=(\log a)_0 \quad \mbox{and} \quad E(a)=\sum_{k=1}^\infty k(\log
a)_k(\log a)_{-k},
\end{equation}
where we define
\begin{equation}
(\log a)_k\equiv\int_{-\pi}^{\pi} \frac{dp}
{2\pi}{\rm e}^{-{\rm i}kp} \log a({\rm e}^{{\rm i}p}).\end{equation}
The theorem also requires some conditions on $a$ to make the infinite sum in $E(a)$ (\ref{GofA}) converge.

If the symbol $a$ has zeros on ${\bf T}$ or has non-zero winding then the infinite sum in $E(a)$ diverges and, therefore, Szeg\"o's theorem cannot be applied.   What can happen instead is that the determinant $\det T_n(a)$ picks up an
algebraic behavior in $n$.  This is described by the Fisher-Hartwig theorem which we introduce next.

We first introduce symbols with a {\it pure Fisher-Hartwig singularity}.
\begin{eqnarray}
\varphi_{\beta,p_0}(\ee^{\ii p})&:=\ee^{\ii\beta(p-p_0)}\quad&\mbox{(pure jump)},
\\\omega_{\alpha,p_0}(\ee^{\ii p})&:=|\ee^{\ii p}-\ee^{\ii p_0}|^{2\alpha}\quad&\mbox{(pure modulus singularity)}.
\end{eqnarray}
Then,
given an arbitrary singular symbol $a$, we factorize $a$ by pure Fisher-Hartwig symbols and a smooth function with zero winding.  Let us restrict ourselves\footnote{A more general case involving multiple Fisher-Hartwig singularities has been considered by Basor and Tracy \cite{BT}.} to one pair of Fisher-Hartwig singularities, and assume $a$ is factorized as $a=\omega_{\alpha,p_0}\varphi_{\beta,p_0}b$ with the conditions that $\alpha,\beta\in{\bf C}$, ${\rm Re}\,\alpha>-1/2$ and $b$ is a smooth function with zero winding.
Then the Fisher-Hartwig theorem states that
\begin{equation}
\det T_n(a)\sim\ee^{nH(b)}n^{\alpha^2-\beta^2}\exp[{E_{\alpha,\beta}(a)}]\quad\mbox{as}\quad n\rightarrow\infty.
\end{equation}
The constant $E_{\alpha,\beta}(a)$ can be found, for instance, in the book \cite{book}.
Here we present the constant for $\alpha=0$, in the form that will be relevant for our calculation.
To do this we define the Hilbert transform of $f(t)=\sum_{j=-\infty}^\infty f_jt^j$ by
\begin{equation}\label{Hilbert1}
f^H(t):=-\frac{1}{\pi}{\cal P}_t\oint d\tau\frac{f(\tau)}{\tau-t}.
\end{equation}
or by
\begin{equation}\label{Hilbert2}
f^H(t):=\frac{1}{\ii}f_0+\frac{1}{\ii}\sum_{j=1}^\infty (f_jt^j-f_{-j}t^{-j}).\end{equation}
In (\ref{Hilbert1}) ${\cal P}_t$ means the principal integral at $\tau=t$.
From now on we will assume the principal integral whenever there is a pole singularity on the integration contour.

Then we can write $E(a)$ in (\ref{GofA}) as
\begin{equation}\label{Eofa}
E(a)=\int_{-\pi}^{\pi}\frac{dp}{4\pi}[(\log a)^H(\ee^{\ii p})]_p\log a(\ee^{\ii p}),\end{equation}
where the subscript $p$ stands for the derivative in $p$, and the branch of the log is chosen to be away from the image of the symbol.
Restricted to the case $\alpha=0$, the constant $E_{0,\beta}(a)$ is given by
\begin{eqnarray}\nonumber\fl
E_{0,\beta}(a)&=E(a/\varphi_\beta)+\ii\beta\left[(\log a/\varphi_\beta)^H(\ee^{\pm\ii\pi})-(\log a/\varphi_\beta)^H_0\right]+\log\left[G(1+\beta)G(1-\beta)\right]
\\\fl&=\int_{-\pi}^{\pi}\frac{dp}{4\pi}\left([(\log a)^H]_p\log a-[(\log\varphi_\beta)^H]_p\log\varphi_\beta\right) +\log\left[G(1+\beta)G(1-\beta)\right],\label{E0beta}
\end{eqnarray}
after a little algebra (see \ref{appendix1}).  Here $(\log a/\varphi_\beta)^H_0$ is the zeroth fourier component of $(\log a/\varphi_\beta)^H$, and $G$ is {\it Barnes G-function} which is an entire function defined by
\begin{eqnarray}
G(z+1)=(2\pi)^{z/2}\rme^{-z(z+1)/2-C_\gamma z^2/2}\prod_{n=1}^\infty\Big\{(1+{\textstyle\frac{z}{n}})^n\,\rme^{-z+z^2/(2n)}\Big\},
\end{eqnarray}
where $C_\gamma\sim0.57721...$ is called the Euler-Mascheroni constant.

\subsection{Eigenvalues distribution from Toeplitz determinants\label{section-theory}}

To study the eigenvalues
$\{\lambda_j|j=1,...,n\}$ of the matrix $T_n(a)$ we consider the determinant,
\begin{equation} \label{TheDet}
\det[\zeta - T_n(a)] = \prod_{i=1}^n (\zeta-\lambda_i).
\end{equation}
It is sufficient to look at this determinant in order to find out about the eigenvalues.

For instance, the log singularities of $\frac{1}{n}\log\det[\zeta-T_n(a)]$ are the places of eigenvalues.  Szeg\"o's theorem tells us that
\begin{equation}\label{assume}
\frac{1}{n}\log\det[\zeta-T_n(a)]\rightarrow H(\zeta-a)=\int_{-\pi}^{\pi}\frac{dp}{2\pi}\log(\zeta-a(\ee^{\ii p})),
\end{equation}
in the limit of infinite $n$.   It is then expected that $H(\zeta-a)$ has a collection of log singularities that coalesce into a line of discontinuity.  The discontinuity, being originated from log singularities, is purely imaginary. This fact unambiguously determines the location of discontinuity along $a({\bf T})$.  As a result the eigenvalues converge to $\{a(\ee^{\ii p})\}$ with the local density of eigenvalues being given by $dp/2\pi$.

\begin{figure}
  \includegraphics[width=8cm]{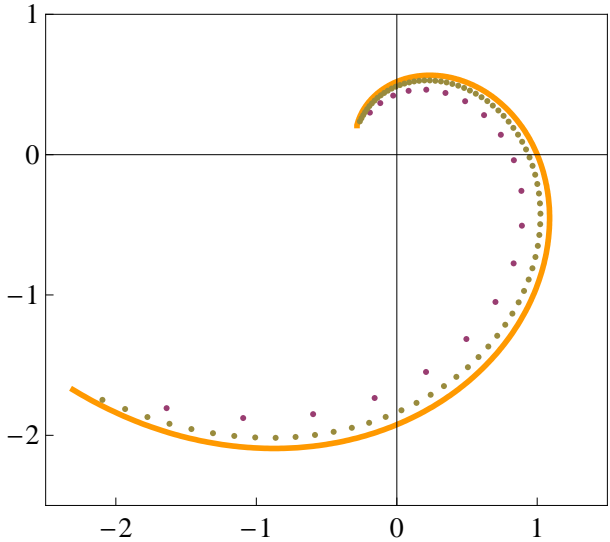}\\
  \caption{The plots of eigenvalues and the image of the symbol in the complex plane.  The horizontal axis showes the real part and the vertical axis showes the imaginary part of the complex coordinate. The eigenvalues are obtained for $T_n(\varphi_\beta)$ where $\beta=\frac{4}{5}+\ii\frac{1}{3}$ and $n=20$ for the inner dots and $n=80$ for the outer dots that are closer to $a({\bf T})$ which is represented by a line.  It shows that the eigenvalues converges to the image of the symbol as $n$ increases.}\label{picture1}
\end{figure}

The above statement is rigorously stated by Widom \cite{Widom} in the following way.
If the convergence (\ref{assume}) holds for almost all $\zeta\in{\bf C}$, then, for an arbitrary continuous function $f$, the following convergence holds.
\begin{equation}
\frac{1}{n}\sum_{j=1}^nf(\lambda_j)\rightarrow\int_{-\pi}^{\pi}\frac{dp}{2\pi}f(a(\ee^{\ii p})).
\end{equation}

Now it is easy to improve the statement (\ref{assume}).
For simplicity, let us assume that $a$ has a single jump singularity and is otherwise continuous.\footnote{We believe that the case with multiple jumps can be similarly considered using the generalized Fisher-Hartwig theorem.}
The Fisher-Hartwig theorem tells us that
\begin{equation}\label{compare2}
\frac{1}{n}\log\det[\zeta-T_n(a)]\sim H(\zeta-a)
-\beta_{\zeta}^2\frac{\log n}{n}+\frac{1}{n}E_{0,\beta_{\zeta}}(\zeta-a),
\end{equation}
where $\beta_\zeta$ characterize the Fisher-Hartwig singularities of the symbol $\zeta-a$ i.e. $\zeta-a$ is factorized into $\varphi_{\beta_\zeta}b_\zeta$ with a smooth, zero-winding function $b_\zeta$.

Previously we argued that the eigenvalues approximate $\{a(\ee^{\ii p})\}$.
By considering the expansion (\ref{compare2})we expect to improve the approximation into $a(\ee^{\ii p})+\delta a(\ee^{\ii p})$.  To obtain the deviation $\delta a$ we simply expand as
\begin{equation}\label{compare1}
\frac{1}{n}\log\det[\zeta-T_n(a)]\sim H(\zeta-a-\delta a)\sim H(\zeta-a)-\int\frac{dp}{2\pi}\frac{\delta a(\ee^{\ii p})}{\zeta-a(\ee^{\ii p})}.
\end{equation}

By comparing (\ref{compare1}) and (\ref{compare2}), and by calculating the discontinuity across the image, we obtain $\delta a$ as
\begin{equation}\label{compareresult}
\delta a(\ee^{\ii\theta})\sim
\ii\partial_\theta a(\ee^{\ii\theta})\left[-\beta_{\zeta}^2\frac{\log n}{n}+\frac{1}{n}E_{0,\beta_{\zeta}}(\zeta-a) \right]^{\zeta=a(\ee^{\ii\theta})+\epsilon}_{\zeta=a(\ee^{\ii\theta})-\epsilon},
\end{equation}
where $\epsilon$ is an infinitesimally small complex number directed normal to the image, i.e. $\propto-\ii\partial_\theta a(\ee^{\ii\theta})$.
In the next subsection, we evaluate the above formula using (\ref{E0beta}).

\subsection{Evaluation of the jump\label{section-theory}}

The goal of this section is to evaluate
$\delta a(\ee^{\ii\theta})$ in (\ref{compareresult}) for an arbitrary continuous symbol $a$ with a single jump.
Without losing generality let us assume that the jump is at $\pm\pi$.

Let us define a few functions.
\begin{equation}\label{realreal}
F_{\pm}(\theta;p):=\log(\zeta-a(\ee^{\ii p}))|_{\zeta=a(\ee^{\ii\theta})\pm\epsilon},
\end{equation}
where $\epsilon$ has already been defined after (\ref{compareresult}) and $F_{\pm}$ are complex-valued functions that coincide in the region $p<\theta$.  $F_{\pm}$ are discontinuous at $p=\theta$ in the following way.
\begin{equation}\label{K}
F_+(\theta;p)+\ii\pi\Theta(p-\theta)=F_-(\theta;p)-\ii\pi\Theta(p-\theta)=:F(\theta;p),
\end{equation}
where $\Theta$ is a step function.  Here we define a continuous function $F$ by the average of $F_{\pm}$.

Evaluating the windings of (\ref{realreal}) in a standard way by
\begin{equation}
\beta_{\pm}(\theta):=\frac{1}{2\pi\ii}\left[F_\pm(\theta;p)\right]_{p=-\pi}^{p=\pi},\end{equation}
the jump of $\beta^2_\zeta$ at $\zeta=a(\ee^{\ii\theta})\pm\epsilon$ is obtained by
\begin{equation}\label{deltabeta2}
\delta\beta^2(\theta):=(\beta_{+}(\theta)+\beta_{-}(\theta))(\beta_{+}(\theta)-\beta_{-}(\theta)) =\frac{\ii}{\pi}\left[F(\theta;p)\right]^{p=\pi}_{p=-\pi},
\end{equation}
where $F$ is defined at (\ref{K}) and we use the fact, $\beta_{+}(\theta)-\beta_{-}(\theta)=-1$.
One also notices that $\beta_{\pm}(\theta)$ is expressed using $\delta\beta^2$ as
$2\beta_{\pm}(\theta)=-\delta\beta^2(\theta)\mp1.$

In the following, we intend to express $\delta a$ of (\ref{compareresult}) using the functions $F$ (\ref{K}) and $\delta\beta^2$ (\ref{deltabeta2}).
Let us evaluate the terms in (\ref{compareresult}) one by one.
First, the $\log n/n$-term is immediately given by $\delta\beta^2$ (\ref{deltabeta2}).

To evaluate $1/n$-term let us recall the integral representation of $E(a)$ (\ref{Eofa}) using the Hilbert transform defined by
\begin{equation}
f^H(x)=\frac{1}{\pi\ii}{\cal P}_x\int_{-\pi}^\pi\frac{\ee^{\ii y}f(y)}{\ee^{\ii y}-\ee^{\ii x}}dy,\end{equation}
for an arbitrary function $f$.

Using the formula (\ref{E0beta}) and $\alpha_\zeta=0$, the jump of $E_{0,\beta_\zeta}(\zeta-a)$ in (\ref{compareresult}) is written:
\begin{eqnarray}\fl\label{big}
\left.E_{0,\beta_\zeta}(\zeta-a)\right|^{\zeta=a(\ee^{\ii\theta})+\epsilon}_{\zeta=a(\ee^{\ii\theta})-\epsilon}
&=\log\left[\frac{G(1+\beta_+(\theta))G(1-\beta_+(\theta))}{G(1+\beta_-(\theta))G(1-\beta_-(\theta))}\right]
\\\nonumber\fl
&+\int_{-\pi}^\pi\frac{dp}{4\pi}\Big[(F_+)^H_pF_+-(F_-)^H_pF_-
\\&\quad-(\log\varphi_{\beta_+(\theta)})^H_p\log\varphi_{\beta_+(\theta)}
+(\log\varphi_{\beta_-(\theta)})^H_p\log\varphi_{\beta_-(\theta)}\Big].
\nonumber\fl
\end{eqnarray}
We first evaluate the terms involving $F_{\pm}$.
Using $F^{+}-F^{-}=-2\pi\ii\Theta(p-\theta)$, we get
\begin{equation}
(F^{+}-F^{-})^H_p=-2\left[\int_\theta^\pi\frac{\ee^{\ii q}\,dq}{\ee^{\ii q}-\ee^{\ii p}}\right]_p
=-\ii\left(\tan\frac{p}{2}+\cot\frac{p-\theta}{2}\right).
\end{equation}
Using this identity we simplify the following terms in (\ref{big}) as
\begin{eqnarray}\label{general}\fl
\int_{-\pi}^\pi\frac{dp}{4\pi}\left[(F_+)^H_pF_+-(F_-)^H_pF_-\right]
&=\int_{-\pi}^\pi\frac{dp}{2\pi}(F_{+}-F_{-})^H_pF
\\\fl&=-\ii\int_{-\pi}^\pi\frac{dp}{2\pi} \left(\tan\frac{p}{2}+\cot\frac{p-\theta}{2}\right)F(\theta;p).
\nonumber
\end{eqnarray}

Similar but simpler calculation simplifies the terms involving $\varphi_{\beta_\pm(\theta)}$ as
\begin{eqnarray}\label{Evarphi}\fl
\int_{-\pi}^\pi\frac{dp}{4\pi}\Big[(\log\varphi_{\beta_+(\theta)})^H_p\log\varphi_{\beta_+(\theta)}
-(\log\varphi_{\beta_-(\theta)})^H_p\log\varphi_{\beta_-(\theta)}\Big]
\\=-\delta\beta^2(\theta)\int_{-\pi}^\pi \frac{dp}{4\pi}\left[\int_{-\pi}^\pi\frac{dq}{\pi\ii}\frac{\ee^{\ii q} q}{\ee^{\ii q}-\ee^{\ii p}}\right]_pp
=-\delta\beta^2(\theta)\int_{-\pi}^\pi\frac{dp}{4\pi}\,p\,\tan\frac{p}{2}\nonumber
\end{eqnarray}

Lastly, we evaluate the term involving the Barnes G-function using the identity $G(z+1)=\Gamma(z)G(z)$ and $\beta_-(\theta)=\beta_+(\theta)+1$ as following.
\begin{eqnarray}\label{logGG}
\log\left[\frac{G(1+\beta_+(\theta))G(1-\beta_+(\theta))}{G(1+\beta_-(\theta))G(1-\beta_-(\theta))}\right] 
=\log\left[\frac{\Gamma(3/2+\delta\beta^2(\theta)/2)}{\Gamma(1/2-\delta\beta^2(\theta)/2)}\right].
\end{eqnarray}

Substituting (\ref{general}), (\ref{Evarphi}) and (\ref{logGG}) into (\ref{compareresult}), we obtain a simplified form of $\delta a(\ee^{\ii\theta})$ as follows.
\begin{eqnarray}\label{final}
\delta a(\ee^{\ii\theta})\sim\ii\partial_\theta a(\ee^{\ii\theta})
\Bigg[-\frac{\log n}{n}\delta\beta^2(\theta)+\frac{1}{n}\times\Omega(\theta)\Bigg].
\\\nonumber\Omega(\theta):=
\int_{-\pi}^\pi\frac{dp}{2\pi}\tan\frac{p}{2}
\Bigg(\frac{\delta\beta^2(\theta)}{2}p-\ii F(\theta;p)\Bigg)
-\ii\int_{-\pi}^\pi\frac{dp}{2\pi}\cot\frac{p-\theta}{2}F(\theta;p)
\\\qquad\qquad+\log\left[\frac{\Gamma(3/2+\delta\beta^2(\theta)/2)}{\Gamma(1/2-\delta\beta^2(\theta)/2)}\right]\nonumber.
\end{eqnarray}
Recall that the eigenvalues approximate $\{a(\ee^{\ii p})$ with the local density of eigenvalues being given by $dp/2\pi$.  We have improved the approximation to $\{a(\ee^{\ii p})+\delta a(\ee^{\ii p})\}$.  Our result (\ref{final}) gives the deviation $\delta a$ for an arbitrary symbol $a$ with a single jump singularity.
In the next section, we demonstrate how this result works by comparing $\delta a(\ee^{\ii p})$ and the deviation of eigenvalues from $a(\ee^{\ii p})$.

\section{Examples and discussion}

\subsection{Numerical comparison}

\begin{figure}
  \subfigure[]{\includegraphics[width=12.5cm]{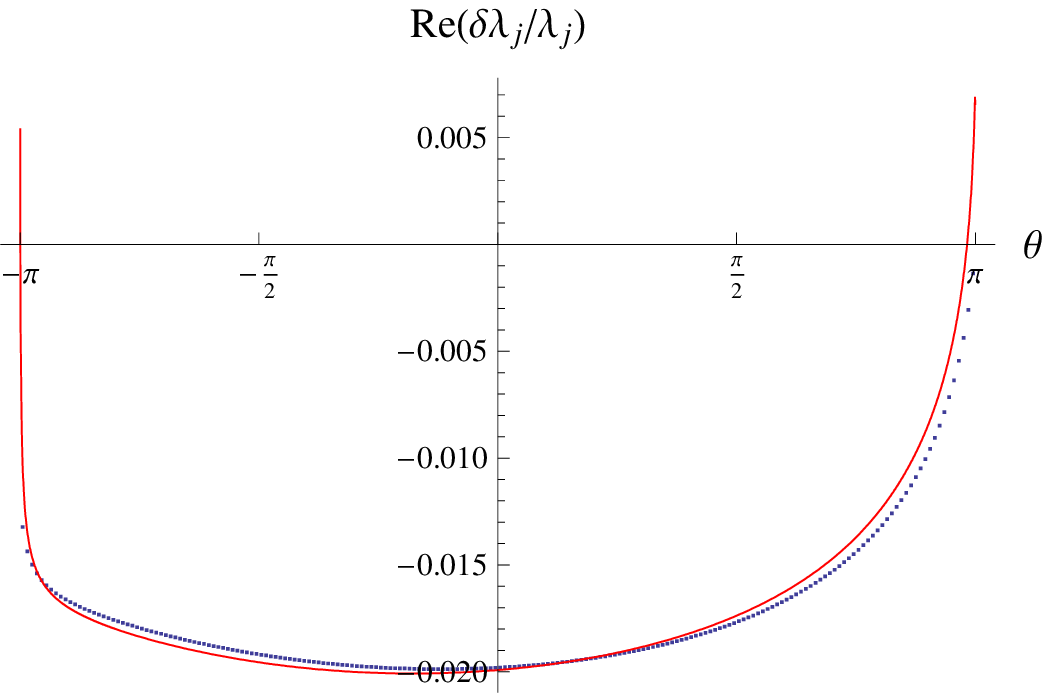}}\\
  \subfigure[]{\includegraphics[width=12.5cm]{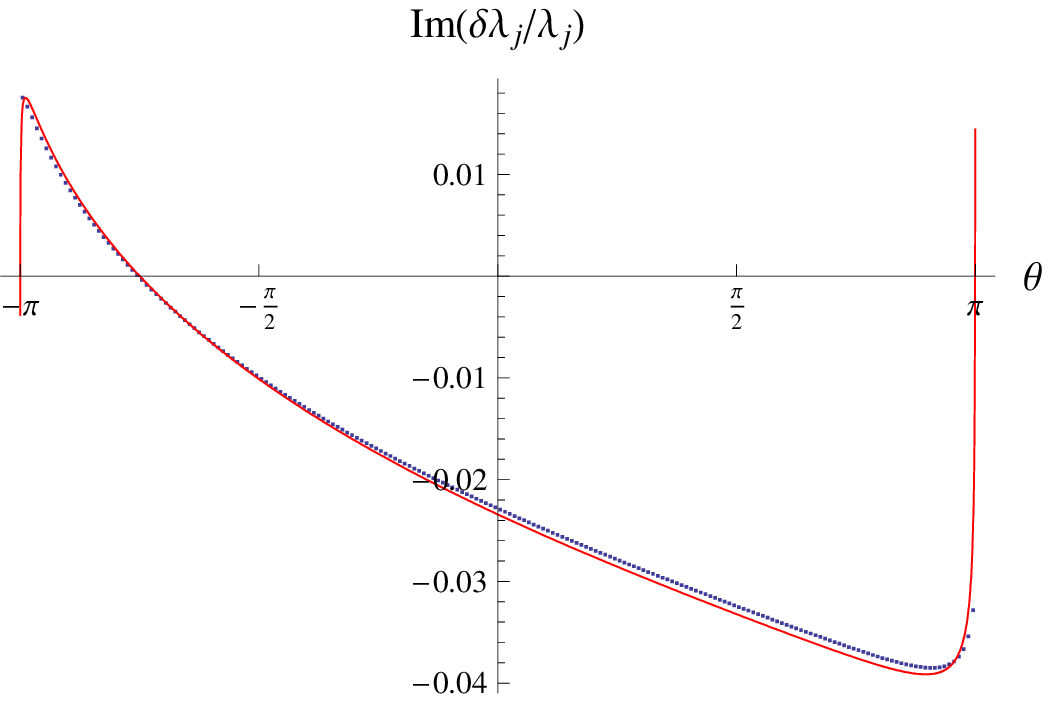}}
  \caption{The deviations of eigenvalues from the image of the symbol and the theoretical predictions. The blue dots are $[\lambda_j-a(\ee^{\ii \theta_j})]/a(\ee^{\ii \theta_j})$ v.s. $\theta_j$.  $\theta_j$ is given by $\frac{2\pi}{n}(j-\frac{1}{2})-\pi$.  The red line is   the theoretical prediction for the blue dots and is given by $\delta a(\ee^{\ii \theta})/a(\ee^{\ii \theta})$ v.s. $\theta$.  The top (a) and the bottom (b) show, respectively, the real and the imaginary part of the plots.  The eigenvalues are obtained for the matrix $T_{n}(\ee^{\ii\beta p})$ where $\beta$ is given by $\beta=\frac{4}{5}+\ii\frac{1}{3}$ and the size is $n=200$.  Though we observed more strong match as we increase $n$ up to 1000 we show the lower $n$ because the dots become indistinguishable at a higher $n$.  One can see from the formula (\ref{final}) that the $y$-axis is of the order $\log n/n$.  Because the $\delta a$ has both $\log n/n$ and $1/n$ components the plots are not scale invariant with $n$.}\label{picture2}
\end{figure}

Here we present an example to illustrate our result (\ref{final}).
We will show that the eigenvalues approximate $\{a(\ee^{\ii \theta})+\delta a(\ee^{\ii \theta})\}$ with the local density of eigenvalues being uniformly distributed in $\theta\in[-\pi,\pi)$.  To do this we choose a most uniformly distributed set $\{\theta_j\}$ by $\theta_j=\frac{2\pi}{n}(j-\frac{1}{2})-\pi$.  And we order the eigenvalues $\lambda_j$'s such that they are close to $a(\ee^{\ii \theta_j})$'s.  This ordering is a straightforward procedure in the example we will present.
To show that the eigenvalues $\{\lambda_j\}$ are approximated by $\{a(\ee^{\ii \theta_j})+\delta a(\ee^{\ii \theta_j})\}$, we plot the actual deviations $\{\lambda_j-a(\ee^{\ii \theta_j})\}$ against the predicted deviations $\{\delta a(\ee^{\ii \theta_j})\}$ to see whether they are close to each other.

We take as our symbol a pure Fisher-Hartwig symbol $a(\ee^{\ii p})=\ee^{\ii\beta p}$ with a general complex $\beta$.
Figure \ref{picture2} shows the plot for $\beta=\frac{4}{5}+\ii\frac{1}{3}$.  The blue dots are the actual deviations and the red line is the theoretical prediction.  They are in good agreement. (See the caption of the figure for more details.)

Being obtained from a perturbative analysis the agreement is of course not perfect.
Also there seems to be an interesting divergence near the ends of the distribution.
We discuss such aspects in the discussion section.

Below we explain how we made these plots.
To get the deviation $\delta a(\ee^{\ii\theta})$ from (\ref{final}) one only has to evaluate the function $F(\theta;p)$ since the other function $\delta\beta^2(\theta)$ is obtained from the former by (\ref{deltabeta2}).
Plugging $\zeta=\ee^{\ii\theta}$ into $\log[\zeta-\ee^{\ii\beta p}]$, and taking the continuous part (i.e. removing the term with $\Theta(p-\theta)$) we get
\begin{eqnarray}\nonumber
F(\theta;p)&= \frac{\ii}{2}\beta(p+\theta+\pi)+\log 2+\ii\arctan\left[
\frac{\cos(\beta_R\frac{\theta-p}{2})\sinh(\beta_I\frac{\theta-p}{2})} {\cosh(\beta_I\frac{\theta-p}{2})\sin(\beta_R\frac{\theta-p}{2})}\right]
\\&\qquad+\frac{1}{2}\log\left(\frac{\cosh\beta_I(\theta-p)-\cos\beta_R(\theta-p)}{2}\right),
\end{eqnarray}
where $\beta=\beta_R+\ii\beta_I$ and we take the standard branch of $\arctan$.

We use the Mathematica to evaluate the integral in (\ref{final}).  Especially, the principal integral at $p=\theta$ is done by the built-in function in the Mathematica, and the other at $p=\pi$ is taken care of by adding an appropriate null integrand, which is proportional to $\tan(p/2)$, to make the integrand to be finite at $p=\pi$.
The numerical evaluation of the eigenvalues is also done with Mathematica.

\subsection{Conclusion and discussion}

It has been observed (and even proven in some cases) that the eigenvalues of certain Toeplitz matrix approximate the image of the symbol.  In this paper we have contributed to the situation by obtaining the leading approximation of the deviations between the eigenvalues and the image of the symbol.

We have derived an explicit formula for an arbitrary symbol with a single jump singularity, and
we have demonstrate our result using a specific example using a pure Fisher-Hartwig symbol.
Since our method is quite general we believe it can be applied to a broader class of symbols.
We believe that the symbols with multiple jumps can be dealt with using the generalized Fisher-Hartwig theorem.
Currently, however, we do not know how to deal with symbols {\it without} jump singularity because we do not know how to evaluate $\det[\zeta-T(a)]$ in large $n$ limit when $\zeta$ is inside the ``closed curve" given by $a({\bf T})$.

Though figure \ref{picture2} shows excellent agreement it is not totally obvious why.  The subtlety lies when we choose $\theta_j$ by the most uniformly distributed points.  The theory tells that the density of eigenvalues are ``uniform" in $\theta$.  But it does not tell whether it is exactly given $\theta_j$'s.  Actually there are many ways to put points with uniform density, for instance, we can move a finite number points while keeping the uniform density in the limit of infinite $n$.  We suspect that this strict uniformity of the eigenvalue distribution seems to require more careful analysis which may even provide a bound to our perturbative analysis.

In figure \ref{picture2} we could observe an interesting structure near the ends of the spectrum. Both the real part and the imaginary part of $\delta a(\ee^{\ii \theta})$ turn out to have a divergence at $\theta=\pm\pi$.
For a real $\beta$, the most dominant structure of the divergence is captured as follows.
\begin{equation}
\frac{\delta a(\ee^{\ii \theta})}{a(\ee^{\ii \theta})}\sim\frac{\ii}{\pi}\left(\frac{\log n}{n}\log(\pi-\theta)+\frac{1}{n}[\log(\pi-\theta)]^2\right),
\end{equation}
near $\theta=\pi$.  From the formula, one notices that the dominant divergences contributes only to the imaginary part, and, quite interestingly, that they exchange their dominance according whether $\log n\gg|\log(\pi-\theta)|$ or $\log n\ll|\log(\pi-\theta)|$.  This competition seems to be making the hook-shaped structure near $\theta=\pi$ in figure \ref{picture2} showing the imaginary part.
Because of these divergence the scaling, $\log n/n$ and $1/n$, of the deviations does not apply to the eigenvalues near the ends of spectrum.  One can expect from figure \ref{picture2} that the eigenvalues near the ends of the spectrum are distributed in a qualitatively different way from those in between.  For instance, we obtain the dominant behavior of $\sim\frac{\log(\log n)\log n}{n}$ for the eigenvalue nearest to the end when we naively apply our result (\ref{final}) to the first eigenvalue.

\subsection*{Acknowledgments}
We want to thank Leo Kadanoff for close supervision and support for this project.  We thank Chris Kempes for his earlier collaboration.  We thank Ilya Gruzberg for the discussions.  SYL thanks Harold Widom and Torsten Ehrhardt for sharing their expertise regarding our results.  The work has been supported in part by the National Science Foundation grant number NSF-DMR 0540811 and by the NSF MRSEC program under NSF-DMR 0213745.  SYL is also supported by ASCII-FLASH.

\appendix
\section{Derivation of (\ref{E0beta})\label{appendix1}}
The derivation goes as follows. The log of $\varphi_\beta$ can be expanded in the following way.
\begin{equation}
\log\varphi_\beta=\beta\left[\log(1+\ee^{\ii p})-\log(1+\ee^{-\ii p})\right]
=-\beta\sum_{k\neq 0}(-1)^{k}\frac{\ee^{\ii k p}}{k}.
\end{equation}
Using this expansion we evaluate the following for an arbitrary function $b$.
\begin{eqnarray}
\int_{-\pi}^{\pi}\frac{dp}{2\pi}[(\log b)^H]_p\log\varphi_\beta&=-\beta\sum_{k\neq 0}\frac{(-1)^{k}}{k}(-\ii k)(\log b^H)_{-k}
\\&=\ii\beta\left[\log b^H(\ee^{\pm\ii\pi})-(\log b^H)_0\right].
\end{eqnarray}
Here the Hilbert transforms are taken not on $b$ but on $\log b$.  One can also check that $\int_{-\pi}^{\pi}\frac{dp}{2\pi}[(\log\varphi_\beta)^H]_p\log b$ gives exactly the same result.
As a result we prove the identity (\ref{E0beta}) in the following way.
\begin{eqnarray}\nonumber\fl
\int_{-\pi}^{\pi}\frac{dp}{4\pi}\left([(\log a)^H]_p\log a-[(\log\varphi_\beta)^H]_p\log\varphi_\beta\right)- E(a/\varphi_\beta)
\\\nonumber=\int_{-\pi}^{\pi}\frac{dp}{4\pi}\left([(\log a)^H]_p\log\varphi_\beta+[(\log\varphi_\beta)^H]_p\log a-2[(\log\varphi_\beta)^H]_p\log\varphi_\beta\right)
\\\nonumber=\int_{-\pi}^{\pi}\frac{dp}{2\pi}\left([(\log a)^H]_p-[(\log\varphi_\beta)^H]_p\right)\log\varphi_\beta
\\\nonumber=\ii\beta\left[(\log a/\varphi_\beta)^H(\ee^{\pm\ii\pi})-(\log a/\varphi_\beta)^H_0\right].
\end{eqnarray}

\end{document}